\documentclass[twocolumns,aps,preprint,superscriptaddress,floatfix,amsmath,amssymb]{revtex4}

\usepackage{graphicx}
\usepackage{dcolumn}
\usepackage{bm}
\usepackage{subfigure}

\begin{document}

\title{Kinetically Inhibited Order in a Diamond-Lattice Antiferromagnet}

\author{G. J. MacDougall}
\email{macdougallgj@ornl.gov}
\affiliation{Neutron Scattering Science Division, Oak Ridge National Laboratory, Oak Ridge, TN, USA, 37831}

\author{D. Gout}
\affiliation{Neutron Scattering Science Division, Oak Ridge National Laboratory, Oak Ridge, TN, USA, 37831}
\affiliation{Materials Science and Technology Division, Oak Ridge National Laboratory, Oak Ridge, TN, USA, 37831}
\affiliation{J$\ddot{u}$lich Centre for Neutron Science-SNS, Forschungszentrum J$\ddot{u}$lich, 52425  J$\ddot{u}$lich, Germany}

\author{J. L. Zarestky}
\affiliation{Ames Laboratory and Department of Physics and Astronomy, Iowa State University, Ames, Iowa 50011, USA}

\author{G. Ehlers}
\affiliation{Neutron Scattering Science Division, Oak Ridge National Laboratory, Oak Ridge, TN, USA, 37831}

\author{A. Podlesnyak}
\affiliation{Neutron Scattering Science Division, Oak Ridge National Laboratory, Oak Ridge, TN, USA, 37831}

\author{M. A. McGuire}
\affiliation{Materials Science and Technology Division, Oak Ridge National Laboratory, Oak Ridge, TN, USA, 37831}

\author{D. Mandrus}
\affiliation{Materials Science and Technology Division, Oak Ridge National Laboratory, Oak Ridge, TN, USA, 37831}
\affiliation{Department of Materials Science and Engineering, University of Tennessee, Knoxville, Tennessee 37996, USA}

\author{S. E. Nagler}
\affiliation{Neutron Scattering Science Division, Oak Ridge National Laboratory, Oak Ridge, TN, USA, 37831}

\begin{abstract} Frustrated magnetic systems exhibit highly degenerate ground states and strong fluctuations, often leading to new physics. An intriguing example of current interest is the antiferromagnet on a diamond lattice, realized physically in A-site spinel materials. This is a prototypical system in three dimensions where frustration arises from competing interactions rather than purely geometric constraints, and theory suggests the possibility of unusual magnetic order at low temperature. Here we present a comprehensive single-crystal neutron scattering study of $\mathrm{CoAl_{2}O_{4}}$, a highly frustrated A-site spinel. We observe strong diffuse scattering that peaks at wavevectors associated with N$\mathrm{\acute{e}}$el ordering. Below the temperature $T^{*}$=6.5 K, there is a dramatic change in the elastic scattering lineshape accompanied by the emergence of well-defined spin-wave excitations. $T^{*}$ had previously been associated with the onset of glassy behavior. Our new results suggest instead that $T^{*}$ signifies a first-order phase transition, but with true long-range order inhibited by the kinetic freezing of domain walls. This scenario might be expected to occur widely in frustrated systems containing first-order phase transitions and is a natural explanation for existing reports of anomalous glassy behavior in other materials.
\end{abstract}

\date{\today}

\maketitle

Frustration occurs in spin systems when constraints prevent the formation of a ground state satisfying all of the pairwise interactions. The defining characteristics of frustration are massive ground state degeneracy and concomitant strong fluctuations. The latter suppress magnetic order and lead to spin-liquid regimes extending to low temperature. There has been great interest in such spin liquids because of the plethora of emergent phenomena\cite{balents10,gardner10,lee10} resulting from an extreme sensitivity to small, often neglected degeneracy-breaking effects. Noteworthy examples include unusual short-range correlations\cite{bergman07,manuel09}, apparent spin-glass behavior in the absence of disorder\cite{gaulin92,greedan96,gardner99}, collective ring excitations in pyrochlore spinels\cite{lee02,kamazawa04,chung05}, spontaneously reduced dimensionality in rare-earth titanates\cite{ross09}, and topological excitations like magnetic monopoles in spin-ice materials\cite{castelnovo08,bramwell09,morris09}.

The degeneracy-breaking effects can be small terms in the interaction Hamiltonian that enter the problem in a non-perturbative way. Examples include dipolar interactions in the pyrochlores\cite{palmer00} or spin-lattice interactions in multiferroics\cite{fabreges09}. Interestingly, degeneracy can also be broken in the complete absence of such interaction terms, through a mechanism known as `order-by-disorder'\cite{villain80}. Order-by-disorder refers to the scenario where the entropy associated with thermal or quantum fluctuations lowers the free energy of one ordered state compared to others of equal energy, thereby stabilizing order. Order-by-disorder was first predicted in studies of the diluted Ising model on a rectangular lattice\cite{villain80}. It has since emerged as a central feature in theories of various frustrated antiferromagnets, including the face-centered cubic (FCC)\cite{henley87,gvozdikova05} as well as some pyrochlore\cite{bramwell94,canals08,champion03,ruff08} and two-dimensional lattices\cite{henley89,chalker92,reimers93}. Despite the long history and ubiquity in theoretical models, few, if any, real materials have been found that exhibit this important phenomenon clearly.

The classical (i.e. thermal-fluctuation induced) order-by-disorder mechanism often results in a first-order phase transition\cite{bergman07,henley87,gvozdikova05,bramwell94}. It is interesting to consider the possible implications of a first-order transition for the low temperature properties of a frustrated magnetic system.  Near and below the transition temperature, T$_{N}$, such a system is necessarily in a limit where the scale of thermal energies is small compared to that of the microscopic magnetic interactions.  Below T$_{N}$, a possible scenario is nucleation at many sites into various near degenerate ground states, resulting in multiple domains separated by antiphase domain walls. For the domains to grow, wall motion must be driven by thermal energy that is sufficient to overcome the magnetic interactions that dominate the surface energy of the walls. Since this condition is not met, the system is kinetically frozen in a state with small domains, and even though it is ordered, it can seem to exhibit glassy behavior regardless of the level of chemical disorder. To our knowledge, this scenario has not been proposed in the literature on frustrated magnets. However, here we present evidence that it occurs in the system $\mathrm{CoAl_{2}O_{4}}$ and, moreover, explains anomalous behavior observed previously in other frustrated materials.

$\mathrm{CoAl_{2}O_{4}}$ is an `A-site spinel' material, where the S=3/2 Co$^{2+}$ ions occupy the tetrahedrally coordinated sites of the spinel structure, forming a diamond lattice. Theoretical investigations\cite{bergman07,lee08,bernier08} have revealed particularly interesting physics in the diamond-lattice antiferromagnets arising from the frustrating effects of competing exchange interactions. The diamond-lattice antiferromagnet with only nearest-neighbor exchange ($J_{1}$) is not frustrated and orders with a collinear ground state as illustrated in Figure~1. However, the lattice has only 4 nearest-neighbors versus 12 next-nearest-neighbors coupled via second neighbor exchange, $J_{2}$. Thus, even moderate $J_{2}$ is predicted to suppress the collinear order, leading to highly degenerate spin-liquid ground states\cite{bergman07} with order-by-disorder type transitions at low temperatures driven by thermal\cite{bergman07,lee08} or quantum fluctuations\cite{bernier08}. In particular, theory predicts that the point $\frac{J_{2}}{J_{1}}=\frac{1}{8}$ is both a tricritical point, where the N$\mathrm{\acute{e}}$el transition switches from second to first-order, and a Lifshitz point, where the low temperature ordered state changes from collinear antiferromagnetism to spin-spiral\cite{bergman07}. In a limited range of $J_{2}/J_{1}$ just above $\frac{1}{8}$, both collinear and spiral spin ordering transitions are predicted, with the upper N$\mathrm{\acute{e}}$el transition driven by the classical order-by-disorder mechanism\cite{bergman07}.

High temperature magnetization measurements show strong antiferromagnetic interactions in $\mathrm{CoAl_{2}O_{4}}$, with a Curie-Weiss temperature around -109 K. Despite this, only glass-like transitions have been reported at a greatly reduced $T^{*}~< $ 9 K\cite{tristan05,suzuki07,tristan08}, implying strong frustration. Powder neutron diffraction has revealed the emergence of short-range spin correlations at low temperatures\cite{krimmel06}, which was reproduced within the spiral-spin-liquid model of Bergman \textit{et al.} by assuming $\frac{J_{2}}{J_{1}}=\frac{1}{8}$. A further analysis of inelastic neutron powder data by Krimmel \textit{et al.}\cite{krimmel09} concluded $\frac{J_{2}}{J_{1}}$ $\sim$0.17 -- well into the regime where spin spiral physics is expected. Contrarily, comparison of diffraction data and Monte Carlo calculations led Zaharko \textit{et al.}\cite{zaharko10} to conclude that $\frac{J_{2}}{J_{1}} < \frac{1}{8}$, suggesting that the system might instead undergo a continuous transition to classical N$\mathrm{\acute{e}}$el order. To date, very little data exists on single crystals, although it has been emphasized on several occasions\cite{bergman07,lee08,krimmel09}that such data is essential to properly compare experiment and theory.

To elucidate the nature of the low temperature state and the significance of $T^{*}$ in $\mathrm{CoAl_{2}O_{4}}$, we have grown single crystals and studied them extensively with neutron scattering. X-ray diffraction showed the crystals to be of high purity with no discernable disorder and minimal Co-Al inversion, refining to $2\%$ with an uncertainty of $4\%$. The magnetization data shows a cusp at $T^{*} \approx$ 6.5 K, in the neighborhood of a broad peak in the heat capacity, consistent with results reported for powders. Characterization of the samples is further discussed in the Supporting Information. Neutron studies of both static magnetic correlations and spin excitations were performed using the High Flux Isotope Reactor (HFIR) and Spallation Neutron Source (SNS) facilities of Oak Ridge National Laboratory (ORNL).

\section{Results and Discussion}
Principal results of elastic scattering measurements are shown in Figure~2. As is observed in $\mathrm{CoAl_{2}O_{4}}$ powders, the most prominent feature in the elastic scattering data is a build-up of intense diffuse scattering at low temperatures. For the regime $\frac {1}{8} < \frac{J_{2}}{J_{1}} < \frac{1}{4}$, theory predicts a spin-spiral state with enhanced diffuse scattering on a spherical surface in the lowest Brillouin zone\cite{bergman07}. What we have observed for $\mathrm{CoAl_{2}O_{4}}$ single crystals is scattering centered about specific Bragg locations with selection rules (H K L) all odd or (H K L) all even with H+K+L = 4n+2 for integer n. This observation implies that the short-range correlations reflect a tendency towards the collinear antiferromagnetic state expected for near-neighbor antiferromagnetic A-site spinels\cite{roth64} (See Figure~1(a).) To explore the correlations in detail we performed radial and transverse scans across several magnetic Bragg positions in the (H H L) scattering plane shown in Figure~1(b). In Figure~2(a), we plot radial scans  across the (0 0 2) Bragg position at three representative temperatures.  The (0 0 2) peak was chosen for illustrative purposes because at this position the structural Bragg peak has vanishing intensity while the magnetic scattering is strongest. A narrow, temperature-independent multiple-scattering peak is visible for T = 25 K and 11 K (see Supporting Information). Magnetic scattering is visible below 100 K and at high temperatures is a single peak that grows in intensity and decreases in width as the temperature is reduced (see e.g. the broad scattering at 11 K in Figure~2(a)). The lineshape changes at low temperatures, as discussed below.

Figure~2(b) shows a map in the (H H L) plane of the elastic scattering at 2 K with that at 25 K subtracted. The signal is clearly centered on the (0 0 2) position and drops off to zero in all directions. The temperature dependence of the magnetic scattering is shown in detail in Figures~2(c) and (d), where the multiple-scattering peak as determined by fits to high temperature data has been subtracted. The logarithmic intensity scale of panel (c) illustrates how the broad scattering at high temperatures narrows significantly and intensifies as the temperature is lowered. The increase in magnetic intensity is most dramatic as T approaches $T^{*}$=6.5 K, as demonstrated by panel(d) showing a series of radial scans plotted on a linear scale for T $<$ 10 K.

At high temperatures, the intensity of the diffuse scattering is well described by a single isotropic Lorentzian function:

\begin{equation}
I(\mathbf{Q},\omega=0) = \frac{A}{(1+\frac{Q^{2}}{\kappa^{2}})},
\label{eq:Lorentz}
\end{equation}

the conventional Ornstein-Zernicke form for short-range correlations above a magnetic phase transition\cite{ornstein14}. Here, $Q$ is the distance from the magnetic Bragg point and the width parameter, $\kappa$, can be associated with the inverse magnetic correlation length. In Figure~3(a), we show the heights and widths extracted from simultaneously fitting radial and transverse scans across (0 0 2) to Eq.~1 up to T=25 K. As the temperature is lowered the strength and extent of the magnetic correlations grow smoothly until saturating at $T^{*}$.  The width of the magnetic diffuse scattering at low temperature is substantially larger than the instrumental resolution width shown as vertical black bars in Figure~2(a).

Below $T^{*}$, there is a qualitative change in lineshape, and a single Lorentzian is no longer sufficient to describe the scattering. This is emphasized by the plot of $\chi^{2}$ in Figure~3(b) showing the failure of the single Lorentzian at low temperatures.  It is further illustrated by the blue line in Figure~2(a), which represents the best fit of a single Lorentzian to the data at T=3.5 K.  In fact the low temperature diffuse scattering cannot be fit well by any physically reasonable single peak lineshape (see details in Supporting Information); below $T^{*}$, a second component of the scattering emerges. The best description of the data is obtained by adding an additional anisotropic Lorentzian-squared function peaked at the same position. Accordingly, the neutron intensities below $T^{*}$ are fit to the two-component form:

\begin{equation}
I(\mathbf{Q},\omega=0) = \frac{A_{1}}{(1+\frac{Q^{2}}{\kappa_{1}^2})}+\frac{A_{2}}{(1+\frac{Q_{||}^{2}}{\kappa_{2,||}^2}+\frac{Q_{\perp}^{2}}{\kappa_{2,\perp}^2})^{2}}.
\label{eq:LplusLsq}
\end{equation}

Here, $Q_{||}$ and $Q_{\perp}$ denote the distance from the magnetic Bragg point, \textbf{G}, of momentum transfer parallel and perpendicular to \textbf{G} in the (H H L) scattering plane, and $\kappa_{2,||}$ and $\kappa_{2,\perp}$ are the respective peak widths in those directions. The quality of the fit for temperatures below 15 K is displayed in Figure~3(b).  At temperatures greater than $T^{*}$ the fit is not improved by adding a second component, however at low temperatures Eq.~2 is a far superior description of the data.

The main panel of Figure~3(c) shows the relative heights of the two components to the scattering. The intensity of the Lorentzian-squared component shows a sudden onset at a temperature that can be identified using linear extrapolation (dashed line) as 6.4$\pm$0.5 K, equal within error to $T^{*}$ identified via bulk probes. The inset shows $\frac{\kappa_{2,||}}{\kappa_{2,\perp}}$, a measure of the anisotropy of the Lorentzian-squared component. In contrast, the Lorentzian component remains isotropic at all temperatures.

The strong temperature dependence and the observed lineshape below $T^{*}$ have important implications for the physics of $\mathrm{CoAl_{2}O_{4}}$. The abrupt emergence of an anisotropic Lorentzian-squared component of the scattering may be a signature of a first-order phase transition at $T^{*}$ to a N$\mathrm{\acute{e}}$el ordered state. This can be understood through the following reasoning. As discussed earlier, below a discontinuous phase transition, ordered regions nucleate in one of the available degenerate ground states and grow until they meet. At this point, the sample volume consists of ordered domains separated by walls. An array of sharp domain walls gives rise to diffuse scattering with a Lorentzian-squared lineshape\cite{debye57}, characteristic of the coarsening regime of first-order phase transition kinetics\cite{nagler88,pouget90}. If there is sufficient thermal energy, the large domains grow and the small ones disappear, resulting in narrower peaks in the diffuse scattering. When the temperature is low compared to the natural energetics of the system, as occurs in frustrated systems, the walls are kinetically frozen and the scattering remains broad. The Lorentzian-squared peak observed below $T^{*}$ in $\mathrm{CoAl_{2}O_{4}}$ suggests antiferromagnetic domains with average size of order 10 spins in each direction. The Lorentzian component of the scattering below $T^{*}$ indicates that some fraction of the spins, most likely those located at walls, remains disordered.

Lorentzian-squared lineshapes have been observed before in frustrated magnets and spin glasses and have usually been attributed to random fields, possibly arising from impurities. We note however that the random field model should not be relevant to a clean single-crystal system with no applied field. Moreover, although random fields can lead to Lorentzian plus Lorentzian-squared (L+L$^{2}$)scattering\cite{birgeneau98}, in principle the $\kappa$ parameters of each component should be equal, and there is no reason to expect anisotropy. In contrast, scattering arising from an array of domain walls is often anisotropic as a consequence of energetics\cite{warren_book,nagler88}. Numerical calculations appropriate for $\mathrm{CoAl_{2}O_{4}}$ (see Supporting Information) show that domain walls perpendicular to the (H H 0) direction have a lower energy than those perpendicular to the (0 0 L) direction. This leads to a smaller average distance between walls and hence broader lineshapes along (H H 0), consistent with experiment.

Given the specific predictions for $\mathrm{CoAl_{2}O_{4}}$\cite{bergman07}, it seems likely that the first-order nature of the transition can be attributed to the classical order-by-disorder mechanism. However, it is worth noting that any discontinuous magnetic transition in a frustrated system might be expected to demonstrate similar behavior. The saturation of the scattering width at value above the resolution width is inconsistent with a continuous transition, where one would expect the correlation length to diverge as the system approaches the transition temperature from above. A direct comparison of the current data with those above a continuous phase transition in a closely related system is provided in the Supporting Information.

Further evidence for the existence of N$\mathrm{\acute{e}}$el order in $\mathrm{CoAl_{2}O_{4}}$ is provided by the appearance of dispersive spin-wave excitations at low temperatures. Figures~4(a,b) show single crystal inelastic neutron scattering spectra taken using the HB1 triple-axis spectrometer at the HFIR for two representative wavevectors, including the magnetic zone boundary (1 0 3).  Well below $T^{*}$ these show clear inelastic modes in addition to a peak centered on zero energy.  As the temperature is raised, the inelastic peaks weaken and broaden as seen in the data at 25 K.  The wavevector dependence of the mode energies in three symmetry directions is depicted in Figures~4(c)-(e) and the mode intensity along the (0 0 L) direction is plotted in Figures~4(f).  The solid red lines in Figures~4(c)-(f) are generated by fitting the available data to a spin-wave model (see Supporting Information) based on the Heisenberg Hamiltonian with nearest (next-nearest) neighbor exchange parameters $J_{1}$ ($J_{2}$) and effective anisotropy field $H_{A}$\cite{lovesey_book}.

The model gives an excellent description of the data, with inferred exchange parameter values $J_{1} = -0.434\pm0.011$~meV and $J_{2} = -0.045\pm0.003$~meV. This implies $\frac{J_{2}}{J_{1}} = 0.104 \pm 0.010$, close to the critical value of $\frac{1}{8}$. The best fits also yield a small but finite value for single-ion anisotropy, $g\mu_{B}H_{A}$ = 0.018 $\pm$ 0.006~meV, implying the existence of a gap in the spectrum.

To resolve the low-energy excitations at the magnetic zone center, cold neutron inelastic scattering measurements were made using the Cold Neutron Chopper Spectrometer instrument at the SNS. The main results are summarized in Figure~5. A representative energy-momentum slice of this data taken at T =1.5 K is plotted in Figure~5(a), with the expected spin-wave prediction superimposed. The adjacent Figure~5(b) is a cut through this data at the magnetic zone center (dashed line). The spectrum at base temperature shows a well-defined magnon peak at the magnetic zone center with a small gap $\Delta = 0.66 \pm 0.04$~meV, in good agreement with extrapolations from the triple axis data. The observation of a well-defined magnon at the zone center argues against a short-range ordered glassy state.  The presence of a gap may be significant, since recent Monte Carlo calculations have suggested that a single-ion anisotropy is necessary to stabilize N$\mathrm{\acute{e}}$el order in $\mathrm{CoAl_{2}O_{4}}$\cite{zaharko10}. The gap likely arises from the effect of spin-orbit coupling which is known to be important for the $Co^{2+}$ ion in a tetrahedral environment\cite{lee08}, and incidentally provides an explanation for the enhanced local moment inferred from bulk magnetization measurements\cite{schlapp32,cossee60}. Finite domain size effects may also play a role.

Figure~5(c), shows an identical data slice at T=8.1 K, just above $T^{*}$. Figure~5(d) directly compares energy cuts for the two temperatures. At 8.1 K the elastic intensity and spin-wave mode easily resolved at low temperature give way to significant quasi-elastic scattering, more typical of a disordered system.

The temperature dependence of the zone center scattering was characterized by fitting the energy cuts to the following form:

\begin{equation}
S(\mathbf{Q},\omega) = A_{1}\cdot G(\omega)+A_{2}\cdot L(\omega)+ A_{3}\cdot DHO(\omega),
\label{eq:tof_cuts}
\end{equation}

where $G(\omega)$ is a Gaussian with width determined by energy resolution describing elastic scattering from the N$\mathrm{\acute{e}}$el-ordered domains, $L(\omega)$ is a Lorentzian parameterizing the quasi-elastic scattering, and $DHO(\omega)$ is a damped harmonic oscillator term describing the magnon.  The functions are normalized to unit area with $L(\omega)$ and $DHO(\omega)$ constructed to satisfy detailed balance. (See Supporting Information for explicit definitions of fitting functions.)  The integrated intensities of the elastic ($A_{1}$) and quasi-elastic ($A_{2}$) components are plotted vs. temperature in Figure~5(e). As expected, the elastic intensity rapidly dies off above $T^{*}$.  The spectral weight of the Lorentzian component increases with temperature up to 8K, weakening slightly at 10.3K. This implies that it is associated with the isotropic Lorentzian component of the Q-dependent elastic scattering.

The data presented here call for a reconsideration of the physical picture describing $\mathrm{CoAl_{2}O_{4}}$, and may have more general implications for additional frustrated magnetic systems. Until now, the behavior of $\mathrm{CoAl_{2}O_{4}}$ below $T^{*}$ has largely been thought of as characteristic of a glassy state, possibly arising from disorder associated with cation site inversion. The single crystals measured here have minimal site inversion, and the current data supports the idea that $T^{*}$ may instead  be associated with a first-order phase transition.  This interpretation naturally explains reports of magnetic order in some powder neutron diffraction\cite{roth64, zaharko10} and ESR\cite{hagiwara10} measurements.  The resultant domain structure is out of equilibrium, consistent with the observed irreversibility in magnetization measurements\cite{tristan05,suzuki07,tristan08}. It should be emphasized that the discontinuities associated with a first-order phase transition become very small on approaching a tricritical point, consistent with the linear temperature dependence of the intensity of the Lorentzian-squared scattering in Figure~3(c), as well as the rounded peak observed in heat capacity measurements. Moreover, heat capacity measurements made with the time relaxation technique have well-known and documented problems handling latent heat\cite{lashley03} and generally will not exhibit of a strong anomaly near a weak first-order phase transition.

The diamond-lattice antiferromagnet with dominant nearest neighbor interactions is expected to exhibit a continuous phase transition to the collinear N$\mathrm{\acute{e}}$el state. In this respect, the first-order nature of the transition is surprising, and in light of the theoretical work of Bergman \textit{et al.}\cite{bergman07}, it is natural to associate it with the physics of frustration and perhaps the order-by-disorder mechanism. It appears that the physics of $\mathrm{CoAl_{2}O_{4}}$ has common features with the closely related FCC Heisenberg antiferromagnet which is also expected from theory to exhibit order-by-disorder physics\cite{henley87,gvozdikova05}. Interestingly, Monte Carlo calculations on the FCC system\cite{gvozdikova05} reveal that domain walls have a significant effect and inhibit the apparent magnitude of the ordered sublattice magnetization. Alternatively, the recent work of Savary \textit{et al.}\cite{savary11} suggests that the ground state properties of certain diamond-lattice antiferromagnets are strongly affected by relatively small amounts of disorder, perhaps enhancing the propensity of $\mathrm{CoAl_{2}O_{4}}$ to form an actual spin glass and leading to strong sample-dependence. We also note certain similarities between the current results and the two length scales observed in the scattering of order-by-disorder candidate $\mathrm{Er_{2}Ti_{2}O_{7}}$\cite{ruff08}, although the latter exhibits a continuous phase transition and resolution-limited Bragg peaks at low temperatures.

It is worth emphasizing again however that the current picture of kinetically inhibited order relies on very few assumptions and may be more generally applicable to systems beyond those which demonstrate order-by-disorder physics. First-order phase transitions are predicted for a number of different reasons including dipolar interactions\cite{melko01} and magneto-elastic coupling\cite{lee00,chung05}, and we believe the picture proposed here for $\mathrm{CoAl_{2}O_{4}}$ is more common than is currently appreciated. The origin of glassy behavior in relatively clean systems is a problem of current interest in the study of rare-earth pyrochlore systems\cite{gaulin92,greedan96,gardner99}. Observations of L+L$^{2}$ lineshapes in elastic neutron scattering exist for some of these systems but remain unexplained\cite{greedan96}. It would very interesting to investigate whether or not there is a universal origin for these features. We believe $\mathrm{CoAl_{2}O_{4}}$ is a prototypical example of an important class of frustrated magnets where apparent glassy behavior is actually a signature of kinetically inhibited magnetic order.


\section*{Materials}
Single crystals were grown via the traveling floating zone method using an NEC image furnace. Quantum Design PPMS and MPMS systems were used to measure the heat capacity and magnetization, respectively. Powder x-ray measurements were performed at the Center for Nanophase Materials Sciences at ORNL.\\
  Triple-axis neutron scattering measurements were performed at the HFIR. Pyrolytic graphite (PG) was used as monochromator and analyzer on both the HB1A and HB1 spectrometers.  Removal of higher order neutron contamination was done via PG filters.  Elastic measurements were carried out on HB1A with a fixed incident neutron energy of 14.6 meV and collimations 48$^{\prime}$-48$^{\prime}$-40$^{\prime}$-68$^{\prime}$ using a single crystal of mass 1.7 g mounted with the (H H L) plane horizontal.  Inelastic measurements were carried out on the HB1 spectrometer using a fixed final energy of either 14.7 meV or 13.5 meV and collimations 48$^{\prime}$-40$^{\prime}$-40$^{\prime}$-120$^{\prime}$.  The sample consisted of the crystal used at HB1A co-aligned within 0.5 degrees with a second crystal for a total mass of 3 g.  The assembly was mounted with the (H H L) and then (H 0 L) scattering planes horizontal in two separate measurements.Time-of-flight measurements used the Cold Neutron Chopper Spectrometer (CNCS) at the SNS.  The specimen was the same single crystal used at HB1A.  CNCS was used with choppers in the `high intensity mode' and the incident energy set to 3 meV for the measurements reported here.

\section*{Acknowledgments}
This Research at Oak Ridge National Laboratory was sponsored by (A) the Materials Science and Engineering Division and (B) the Scientific User Facilities Division, Office of Basic Energy Sciences, U. S. Department of Energy. The authors would like to acknowledge T. Hong, A.D. Christianson and J.L. Niedziela for technical assistance during neutron scattering measurements and B.C. Sales and A.S. Sefat for support during the preparation and characterization of single crystals. The authors would also like to acknowledge B.C. Larson for a critical reading of the manuscript, and M.J.P. Gingras for critical comments and useful discussion.

\clearpage
\section{Figure Legends}
\vspace{10mm}

\noindent
\textbf{Figure 1.}

\textbf{Structure of $\mathrm{CoAl_{2}O_{4}}$.} ~(\textbf{A}) Magnetic cobalt ions of $\mathrm{CoAl_{2}O_{4}}$ on the A-site diamond-lattice. This can be deconstructed into two interpenetrating FCC sublattices (colored here blue and orange, respectively). Arrows represent the ordered state observed to exist in diamond-lattice antiferromagnets when nearest-neighbor exchange dominates. (\textbf{B}) The (H H L) plane of reciprocal space for the diamond-structure using cubic notation. Blue diamonds denote allowed structural Bragg peaks. Red circles denote allowed magnetic Bragg peaks for the ordered state shown in (\textbf{A}). At reciprocal points indexed by three odd integers, both nuclear and magnetic Bragg scattering is expected. Green lines illustrate directions in reciprocal space for which spin-wave dispersions are plotted in Figure~4.

\vspace{10mm}
\noindent
\textbf{Figure 2.}

\textbf{Short-range magnetic correlations at low temperatures.}~(\textbf{A}) Radial scans across the (0 0 2) position plotted for three temperatures.  The intensity is normalized to a fixed number of incident neutrons equaling approximately 10 seconds of counting time. Error bars are $1\sigma$ from Poisson counting statistics. Strong diffuse scattering emerges at low temperature. Solid curves through the data points represent lines of best fit to Eq.~2 plus an additional narrow Gaussian term to account for multiple scattering. The solid blue line shows a fit of Eq.~1 to the $T = 3.5$ K data. Vertical black bars (bottom) show the full-width of the instrumental resolution in the radial direction.  ~(\textbf{B}) The scattering intensity at T = 2 K minus that at T = 25 K in an extended region of reciprocal space about the (0 0 2) Bragg position. The diffuse scattering is clearly centered on the Bragg position and quickly goes to zero in every direction in reciprocal space. Intensity is presented on a logarithmic scale. ~(\textbf{C}) Scattering intensity along the (0 0 L) direction as a function of temperature up to 25 K. Intensity is plotted on a logarithmic scale. ~(\textbf{D}) A series of radial scans across the (0 0 2) position, for temperatures below 10 K and with intensity on a linear scale. In both (\textbf{C}) and (\textbf{D}), the small peak due to multiple scattering has been subtracted. Solid curves in (\textbf{D}) are lines of best fit to Eq.~2.

\vspace{10mm}
\noindent
\textbf{Figure 3.}

\textbf{Emergent component to the scattering profile.} Combined data for radial and transverse scans across the (0 0 2) peak were fit to Eq.~1 and Eq.~2. ~(\textbf{A}) Peak heights and widths extracted from fits to the single peak Lorentzian form Eq.~1.   ~(\textbf{B}) Comparison of the normalized $\chi^{2}$ for each fit as a function of temperature.
~(\textbf{C}) Peak heights extracted using two-peak form Eq.~2 to fit data below T=$T^{*}$. Blue squares denote the height of the Lorentzian component, red circles that of the Lorentzian squared component. ~\textbf{Inset} Ratio of Lorentzian squared width in the 00L direction to that in the HH0 direction.  Error bars in all plots are statistical $1\sigma$ extracted from least-square fits.

\vspace{10mm}
\noindent
\textbf{Figure 4.}

\textbf{Collective spin-wave excitations.} ~(\textbf{A,B}) Representative neutron scattering spectra at (H K L)=(0.4 0 2.4) and (1 0 3), respectively, at temperatures of 1.5 K and 25 K. Counts are normalized to a fixed number of incident neutrons corresponding to approximately 2 minutes per point.  Error bars are $1\sigma$ assuming Poisson counting statistics.  ~(\textbf{C-E}) Excitation energies extracted from fits to individual scans plotted vs. wavevector passing through the (0 0 2) point along high-symmetry directions (H 0 H) (C), (H H H) (D), and (0 0 L) (E). Solid curves show the dispersion predicted by fitting all data to linear spin-wave theory for the ordered state shown in Figure~1(a). ~(\textbf{F}) Intensity of the spin-wave mode along the (0 0 L) direction.  Solid line is the prediction of linear spin-wave theory. Error bars in ~\textbf{C-F} correspond to statistical errors from fits.

\vspace{10mm}
\noindent
\textbf{Figure 5.}

\textbf{Zone center excitations and anisotropy gap.} ~(\textbf{A}) Neutron scattering intensity near the magnetic zone center (0 0 2) at T = 1.5 K, plotted as a function of energy and momentum along (H 0 H). Color bar shows the intensity in arbitrary units. The red line represents the expected dispersion curve for spin-waves in a N$\mathrm{\acute{e}}$el antiferromagnet, using parameters extracted from data in Figure~4.   ~(\textbf{B}) Energy cut through the same data along the dashed line in (A). Error bars are $1\sigma$ from counting statistics. The solid line shows fit to Eq.~3. ~(\textbf{C}) As panel (A) but  for T = 8.1 K.  ~(\textbf{D}) Cuts as in panel (B), plotted on a logarithmic intensity scale for T = 1.5 K and T = 8.1 K. Solid lines represent fits to Eq.~3.  ~(\textbf{E}) Integrated intensities of the elastic (Gaussian, $A_{1}$) and quasi-elastic (Lorentzian, $A_{2}$) components derived from fits of data to Eq.~3. The error bars are $1\sigma$ uncertainties in the fitted parameters, and the solid lines are guides to the eye.

\clearpage
\section*{Figures}

\begin{figure}[h]
\centering
\includegraphics[width=\columnwidth]{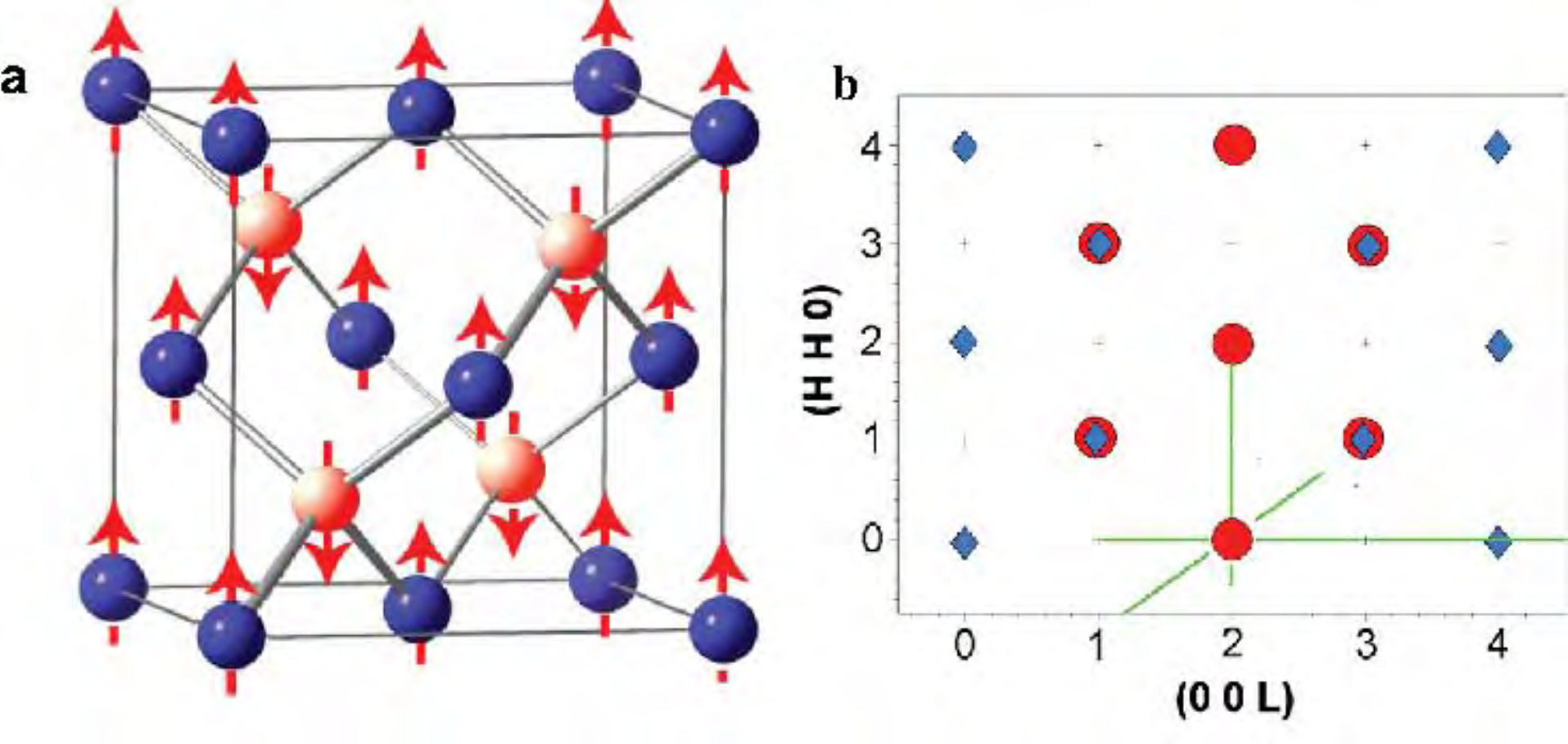}
\caption{\label{fig:structure} }
\end{figure}

\begin{figure}[tpb]
\centering
\includegraphics[width=\columnwidth]{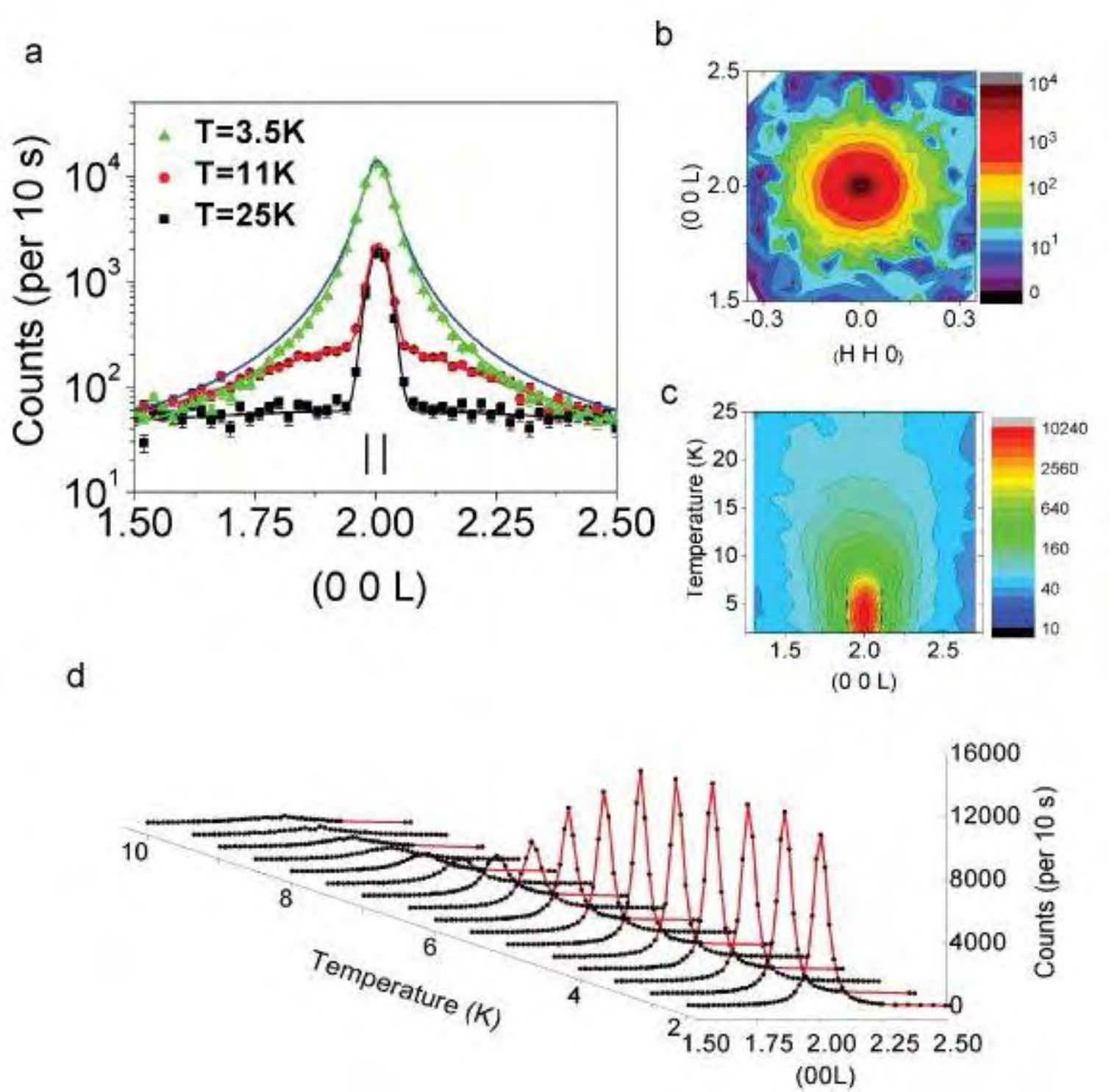}
\caption{\label{fig:TA_diffuse} }
\end{figure}

\begin{figure}[tpb]
\centering
\includegraphics[width=0.5\columnwidth]{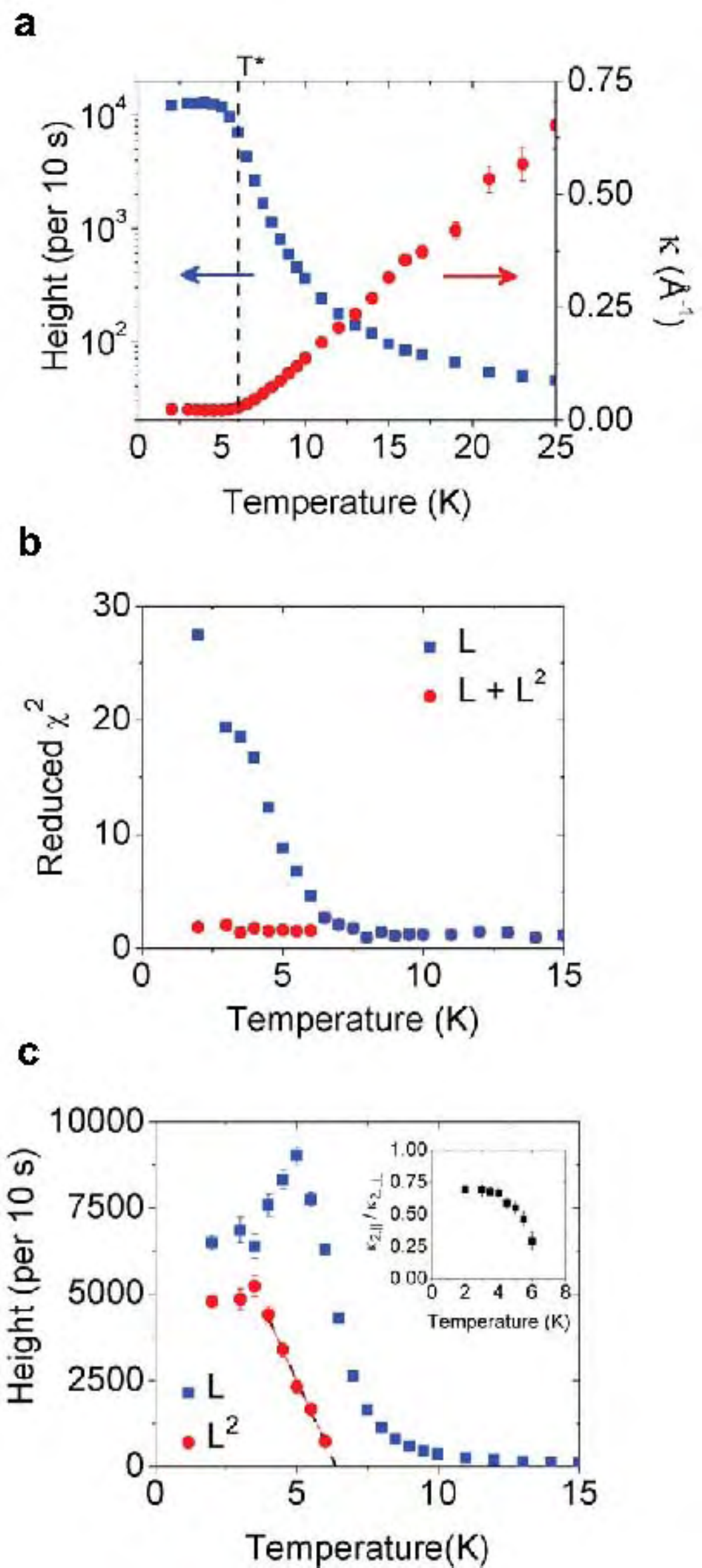}
\caption{\label{fig:TA_fits} }
\end{figure}

\begin{figure}[tpb]
\centering
\includegraphics[width=\columnwidth]{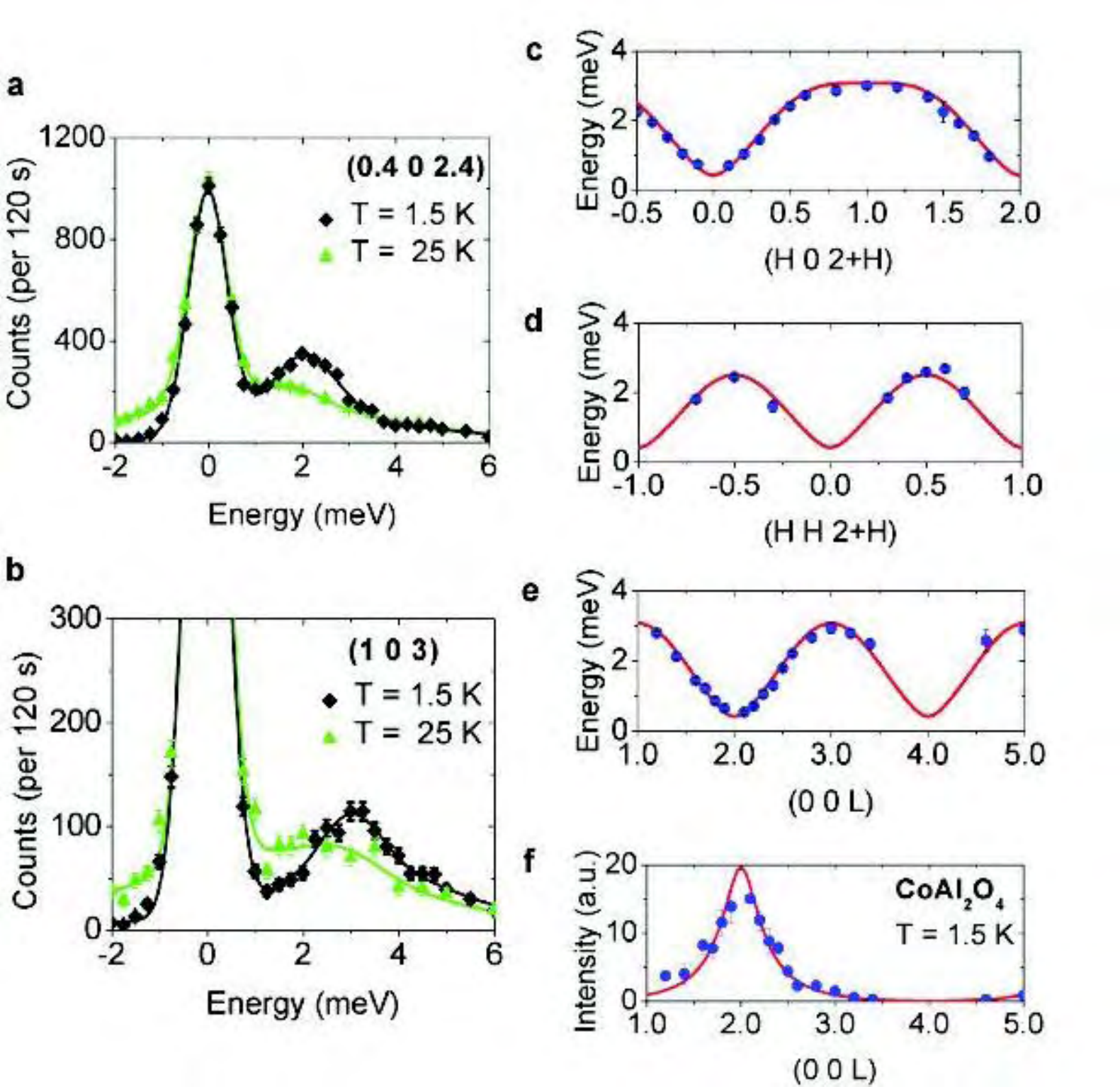}
\caption{\label{fig:TA_inelastic} }
\end{figure}

\begin{figure}[tpb]
\centering
\includegraphics[width=\columnwidth]{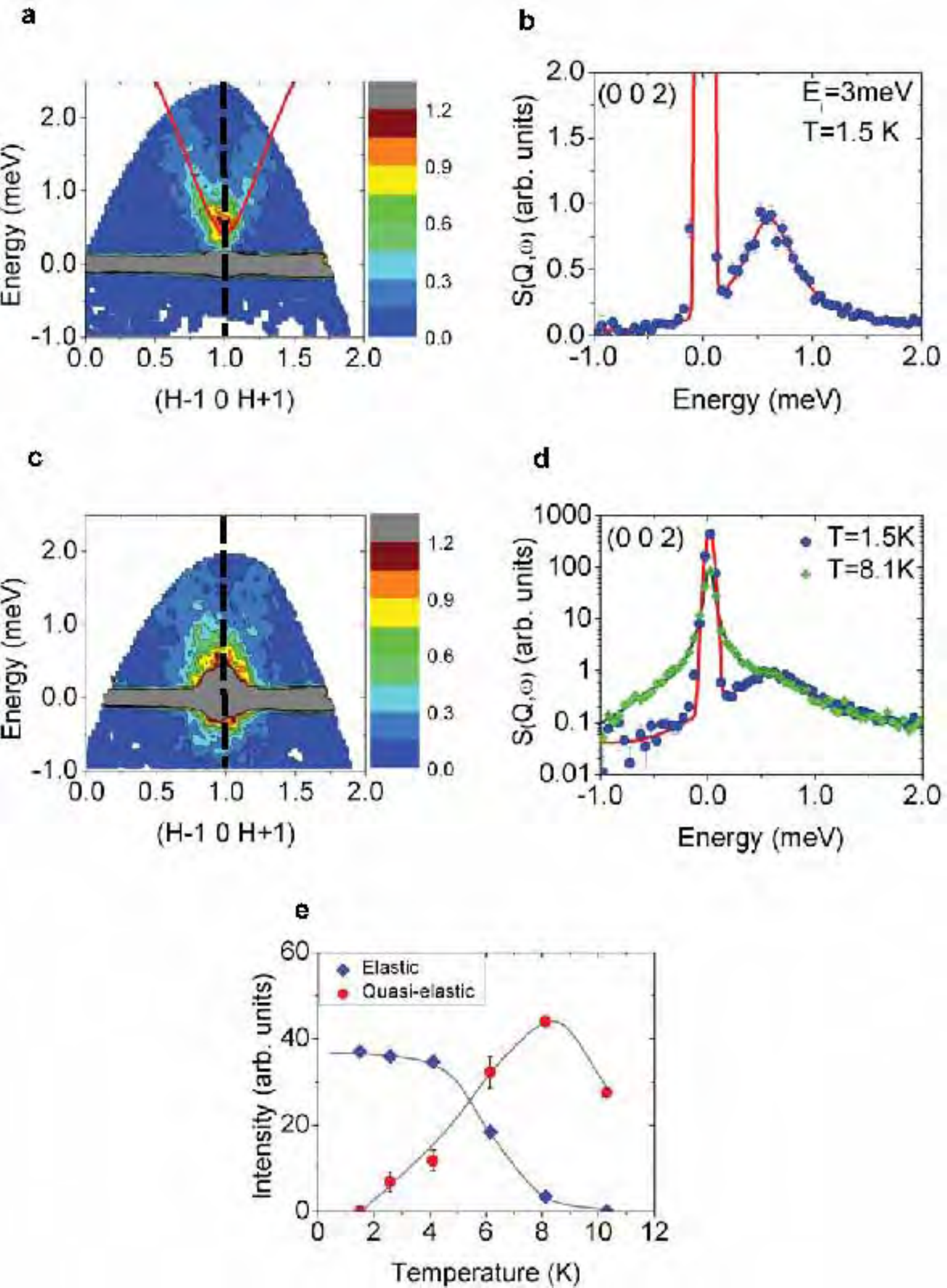}
\caption{\label{fig:TOF}}
\end{figure}

\clearpage
\setcounter{figure}{0}
\setcounter{equation}{0}
\setcounter{section}{0}

\begin{center}
\large
\textbf{SUPPORTING INFORMATION}
\end{center}

\
\section{Crystal Growth and Characterization}

Crystals for this study were grown using an NEC optical image furnace, with growth rates varying between 10 mm/hr and 15 mm/hr and in a flowing $O_{2}$ environment. The last mm of growth was cut from one crystal, ground into powder and characterized with powder x-ray diffraction and SQUID magnetization. Heat capacity measurements were made with a separate slice of crystal.

The x-ray pattern verified the spinel crystal structure expected for $\mathrm{CoAl_{2}O_{4}}$, with no discernable impurity phases. Refinement gave a crystal lattice parameter of 8.107 \AA~ at room temperature, consistent with values reported by Tristan \em{et al.}~\em\cite{tristan05,tristan08}.  The inversion parameter, x=0.02$\pm$0.04 is among the lowest reported for this material.

Bulk magnetization was measured with a Quantum Design MPMS SQUID magnetometer. The results measured in an applied field of 1 T are shown in Figure~S1(a). The high temperature data show Curie-Weiss behavior, which can be fitted to give an effective moment of $p_{eff}$=$4.89\pm0.03~ \mu_{B}$ and Curie-Weiss constant $\Theta$=-109$\pm$1 K, again consistent with values reported in the literature\cite{tristan05,tristan08}. The fitted effective moment is larger than expected for a spin-only system with S=3/2. This is a well-understood effect in tetrahedrally-coordinated $Co^{2+}$ and is principally arising from spin-orbit coupling. (See, for example, Refs.~\cite{schlapp32,cossee60}.) The inset in Figure~S1(a) shows a closeup of the cusp in the magnetization, identifying $T^{*}\approx$6 K. The low temperature field dependence of the bulk magnetization was measured up to 5 T and found to be perfectly linear with field.  Non-linearity with field has been observed in the past to be associated with a $\mathrm{Co_{3}O_{4}}$ impurity\cite{tristan05}. Such an impurity can be ruled out to within the sensitivity of our measurements.

Heat capacity was measured with a Quantum Design PPMS using the relaxation method: the extracted data are shown in Figure~S1(b). The temperature dependence shows a broad peak at T slightly above $T^{*}$, consistent with previous published results on polycrystalline material\cite{tristan05,tristan08,suzuki07}. Well below $T^{*}$, specific heat varies with a $T^{2}$ temperature dependence. This is consistent with the results of Tristan \textit{et al.} \cite{tristan05,tristan08}, but different from that measured by Suzuki \textit{et al.} \cite{suzuki07} and that predicted by Bergman \textit{et al.} for a spiral spin ordered ground state\cite{bergman07}. It is also contrary to the linear specific heat dependence expected for a canonical spin glass system.

The crystals were characterized via neutron scattering, as laid out in the main article. The Bragg peaks expected for the ideal $\mathrm{CoAl_{2}O_{4}}$ spinel structure were observed.  Inversion between $Co^{2+}$ and $Al^{3+}$ cations or vacancies at atomic sites would lead to additional elastic scattering at nominally forbidden peak positions such as (0 0 1), (0 0 3), (1 1 0) and (3 3 0).   No intensity above background was observed at any of these positions, consistent with the low level of inversion implied by the powder x-ray measurement.  A representative `rocking curve' for the crystal used in the elastic neutron scattering measurements is shown in Figure~S2 for the nuclear Bragg peak at (H K L) = (0 0 4). The observed crystal mosaic was $\sim$30$^{\prime}$, and other crystals had a similar or smaller mosaic spread. These values are comparable to those reported for crystals grown elsewhere by the same method\cite{madjuk09}.

\section{Multiple Scattering}

The wavevector (0 0 2) is not an allowed Bragg peak in the spinel structure. The appearance of a small peak at (0 0 2) in our data at high temperatures (Fig.~1 of the main article) is understood to result from multiple scattering\cite{shirane_book}. The elastic scattering data presented in the main article was taken with the crystal oriented in the (H H L) scattering plane, giving the possibility of an apparent signal at (0 0 2) arising from double scattering events involving pairs of allowed Bragg peaks, for example (1 1 3) and (-1 -1 -1).

Multiple scattering can often be distinguished from direct scattering because the former is highly dependent on the scattering plane and its intensity can change drastically as the sample is rotated about the scattering wavevector.  Figure~S3 shows that the intensity at (0 0 2) is drastically different depending upon whether the sample is  oriented in (H 0 L) scattering plane versus (H H L). The small amount of signal seen at (0 0 2) in the (H 0 L) plane is consistent with expectations for scattering from the (0 0 4) Bragg peak by $\lambda$/2 neutrons that were not perfectly removed by the graphite filters.  The larger signal at (H H L) is completely consistent with multiple scattering.

The multiple scattering signal at (0 0 2) shows very little temperature dependence below 100 K, and apparently none at all below 25 K.  Above 10 K it is easily distinguished from the magnetic signal, and it has been modeled in the low temperature data using the fit to the 10 K data. At 2 K it contributes a small fraction of the total intensity at the precise peak position, and no intensity whatsoever to the wings of the magnetic scattering.   The amount of magnetic scattering at the allowed Bragg peaks that contribute to the multiple scattering intensity at (0 0 2) is sufficiently small that it does not affect the data analysis.  In fact, it has been verified that the conclusions drawn from the data analysis at low temperatures are not strongly affected even if the fitting procedure simply ignores the multiple scattering contribution.

\section{Lineshape Analysis}

As mentioned in the main article, magnetic scattering at the (0 0 2) reciprocal lattice position was well-described by an isotropic Lorentzian function at temperatures greater than $T^{*}$. Upon cooling through this temperature though, a sudden change in the lineshape was observed. To explore the form of the scattering function at low temperatures, radial and transverse scans across the (0 0 2) reciprocal lattice position were fit simultaneously to a number of different functions. Table~S1 lists the normalized $\chi^{2}$ for several of the attempted fits of the data at T=2 K and T=3K. Included are results of fits to different combinations Gaussian (G), Lorentzian (L), square Lorentzian ($L^{2}$), and Lorentzian to an arbitrary power ($L^{p}$). Each function was either constrained to be isotropic (I) are allowed to be anisotropic (A).

As is clear from the quality-of-fit parameters, no single-component function describes the data well at lowest temperature, nor does any perfectly isotropic function. In fact, by a large margin, the peak functions described by a Lorentzian and an anisotropic Lorentzian-squared function give the best description of the data. In these fits, the anisotropy ratio of the width parameters for the Lorentzian-squared component, $\frac{\kappa_{2,\|}}{\kappa_{2,\perp}}$, ranges from 0.3 to 0.7.  If one allows the width parameters describing the Lorentzian component to vary independently in the radial and transverse directions, with the single exception of the data at T=2K the resulting best fits yield a width parameter equal to 1 within 3\% and no improvement in the quality of fit as measured by the normalized $\chi^{2}$.  The data at 2 K exhibits a difference of 10\%.

It is important to note that the conclusions reached here are unaffected by the inclusion of instrument resolution in the fits. As mentioned and shown graphically in Figure~2 of the main article, the width of the elastic scattering peak is significantly larger than the resolution widths. It has been verified that the $IL+AL^{2}$ form remains the most accurate description of the data when the effects of resolution are included. Moreover, inferred signal anisotropy of the Lorentzian-squared term is in fact accentuated by the inclusion of resolution, as the resolution ellipsoid is considerably broader in the (0 0 L) than the (H H 0) direction at the (0 0 2) reciprocal lattice position.

\section{Domain Walls}

As argued in the main article, the anisotropy of the diffuse scattering widths in $\mathrm{CoAl_{2}O_{4}}$ at lowest temperatures can be naturally interpreted as a signature of preferred wall orientation in a N$\mathrm{\acute{e}}$el ordered state with domains. Specifically, the broader observed linewidth in the (1 1 0) relative to the (0 0 1) crystallographic direction implies a shorter distance between domain walls in that direction.

To follow up on this idea, we numerically estimated the number of nearest and next-nearest neighbor bonds broken by an anti-phase domain wall of each orientation in an N$\times$N$\times$N array of cubic cells containing the N$\mathrm{\acute{e}}$el order for different values of N. We then extrapolated these estimations to N$\to \infty$. The energy of a wall was then derived from the total energy cost (gain) of breaking nearest-neighbor (next-nearest-neighbor) bonds via:
\begin{equation}
E_{wall} = 2|J_{1}|\times N_{broken,1} - 2|J_{2}|\times N_{broken,2}.
\end{equation}
Here, we used the values of $J_{1}$ and $J_{2}$ extracted from fits of inelastic data to linear spin-wave theory. Results for walls oriented normal to the (0 0 1) and (1 1 0) directions are listed in Table~S2.

Consistent with experiment, walls oriented normal to the (0 0 1) direction have a higher estimated energy cost per unit area, implying a larger distance between such walls and a narrower neutron scattering line-width in that direction.

\section{Saturation of Magnetic Correlation Length}

At temperatures greater than the magnetic transition temperature, the current study observes intense Lorentzian-like elastic scattering which grows in intensity and correlation length (inversely proportional to peak width) as $T_{N}$ is approached from above. Importantly though, the magnetic correlation lengths are seen to saturate at a value of $\kappa\approx$33\AA, which is much shorter than the resolution length of the instrument. This is in direct contrast to the behavior above a continuous phase transition to long-range order, where correlation lengths are seen to diverge.

To emphasize this point, we plot in Figure~S4 the fitted values of $\kappa$ from the isotropic Lorentzian component of the scattering in $\mathrm{CoAl_{2}O_{4}}$ and compare to equivalent data from a separate study of $\mathrm{MnAl_{2}O_{4}}$. $\mathrm{MnAl_{2}O_{4}}$ is also a diamond-lattice antiferromagnet, but has a smaller $J_{2}$ and a continuous phase transition to long-range N$\mathrm{\acute{e}}$el order at $T_{N}$=39 K\cite{tristan05}. Notably, the $\mathrm{MnAl_{2}O_{4}}$ data shown here are from a sample with a larger level of cation inversion, and thus greater amount of disorder, than the current crystal of $\mathrm{CoAl_{2}O_{4}}$. Nevertheless, the extent of Ornstein-Zernicke correlations at the transition temperature is much longer in the $Mn$ material. Comparisons such as this provide further evidence that a continuous transition is avoided in $\mathrm{CoAl_{2}O_{4}}$.

\section{Spin-waves on a Diamond Lattice}

The linear spin-wave predictions used to fit inelastic data in the main text were based on the derivations of Lovesey\cite{lovesey_book} and are laid out below.
Consider the following Hamiltonian for an antiferromagnet described by two interpenetrating lattices, \textbf{m} and \textbf{n}, containing spin-up and spin-down moments, respectively:
\begin{eqnarray}
\mathcal{H} &=& J_{1}\sum_{\bm{m},\bm{r}}\bm{S}_{\bm{m}}\cdot \bm{S}_{\bm{m}+\bm{r}}+J_{1}\sum_{\bm{n},\bm{r}}\bm{S}_{\bm{n}}\cdot \bm{S}_{\bm{n}+\bm{r}}+ J_{2}\sum_{\bm{m},\bm{r}^{'}}\bm{S}_{\bm{m}}\cdot \bm{S}_{\bm{m}+\bm{r}^{'}}\\ && + J_{2}\sum_{\bm{n},\bm{r}^{'}}\bm{S}_{\bm{n}}\cdot \bm{S}_{\bm{n}+\bm{r}^{'}}
-g\mu_{B}H_{A}\sum_{\bm{m}}\bm{S}_{\bm{m}}^{z} + g\mu_{B}H_{A}\sum_{\bm{n}}\bm{S}_{\bm{n}}^{z}.
\end{eqnarray}

Here, \textbf{r} is defined to be any vector which connects nearest-neighbors (coupling opposite sublattices), \textbf{r}$^{'}$ is any vector which connects next-nearest neighbors (n.n. on a single sublattice) and $H_{A}$ is an effective field along the (arbitrarily chosen) \textbf{z}-axis included to parameterize a possible uniaxial anisotropy. Potential sources of exchange anisotropy in A-site spinels have been discussed at length by Lee and Balents\cite{lee08}. One candidate is spin-orbit coupling in the form of a Dzyaloshinskii-Moriya interaction, allowed here since the diamond lattice breaks bond inversion symmetry. As mentioned previously, spin-orbit coupling has long been known to be important in systems with $Co^{2+}$ ions in a tetrahedral environment and can explain the enhanced local moment inferred from high-temperature magnetization measurements\cite{schlapp32,cossee60}. An alternate potential origin is the dipole-dipole interaction between $Co^{2+}$ spins.

For a diamond lattice, there are 4 nearest neighbors separated by the vectors\\
    $\mathbf{r}_{1} = [\frac{a}{4}~ \frac{a}{4}~ \frac{a}{4}]$\\
    $\mathbf{r}_{2} = [\frac{-a}{4}~ \frac{-a}{4}~ \frac{a}{4}]$\\
    $\mathbf{r}_{3} = [\frac{-a}{4}~ \frac{a}{4}~ \frac{-a}{4}]$\\
    $\mathbf{r}_{4} = [\frac{a}{4}~ \frac{-a}{4}~ \frac{-a}{4}]$\\

and 12 next-nearest neighbors separated by\\
    $\mathbf{r}^{'}_{1} = [\frac{a}{2}~ \frac{a}{2}~ 0]$\\
    $\mathbf{r}^{'}_{2} = [\frac{a}{2}~ \frac{-a}{2}~ 0]$\\
    $\mathbf{r}^{'}_{3} = [\frac{-a}{2}~ \frac{a}{2}~ 0]$\\
    $\mathbf{r}^{'}_{4} = [\frac{-a}{2}~ \frac{-a}{2}~ 0]$\\
    $\mathbf{r}^{'}_{5} = [\frac{a}{2}~ 0~ \frac{a}{2}]$\\
    $\mathbf{r}^{'}_{6} = [\frac{a}{2}~ 0~ \frac{-a}{2}]$\\
    $\mathbf{r}^{'}_{7} = [\frac{-a}{2}~ 0~ \frac{a}{2}]$\\
    $\mathbf{r}^{'}_{8} = [\frac{-a}{2}~ 0~ \frac{-a}{2}]$\\
    $\mathbf{r}^{'}_{9} = [0~ \frac{a}{2}~ \frac{a}{2}]$\\
    $\mathbf{r}^{'}_{10} = [0~ \frac{a}{2}~ \frac{-a}{2}]$\\
    $\mathbf{r}^{'}_{11} = [0~ \frac{-a}{2}~ \frac{a}{2}]$\\
    $\mathbf{r}^{'}_{12} = [0~ \frac{-a}{2}~ \frac{-a}{2}]$\\

Using this information, one can define:\\
\begin{align}
\mathcal{J}(\mathbf{Q}) &= \sum_{n.n.} J_{1}~ e^{i\mathbf{Q}\cdot r_{j}}\nonumber\\
             & = 4J_{1}~ \{\cos(\frac{\pi h}{2})\cos(\frac{\pi k}{2})\cos(\frac{\pi l}{2}) - i \sin(\frac{\pi h}{2})\sin(\frac{\pi k}{2})\sin(\frac{\pi l}{2})\}
\label{eq:Jprime}
\end{align}
to describe interactions between spins on opposite sublattices, and
\begin{align}
\mathcal{J}^{'}(\mathbf{Q}) &= \sum_{n.n.n.} J_{2}~ e^{i\mathbf{Q}\cdot r^{'}_{j}}\nonumber\\
             &= 2J_{2}~ \{ \cos(\pi (h + k)) + \cos(\pi (h - k)) + \cos(\pi (h + l))\nonumber\\
             & + \cos(\pi (h - l)) + \cos(\pi (k + l)) + \cos(\pi (k - l))\}
\label{eq:J}
\end{align}
to describe interactions between spins on same sublattice. Above, we continue to consider only nearest and next-nearest neighbor interactions on the diamond lattice, and we have used the convention $\mathbf{Q} \equiv \dfrac{2\pi}{a}(h~ k~ l)$. It is interesting to note that the function $\mathcal{J}(\mathbf{Q})$ is a complex quantity. This is a direct result of the diamond lattice being non-Bravais and the sites of one sublattice not being centers of symmetry with respect to sites of the opposite sublattice.

In this situation, linear spin-wave theory predicts that the magnon dispersion is given by\cite{lovesey_book}:\\
\begin{equation}
\omega(\mathbf{Q}) = 2S\sqrt{(\mathcal{J}(0) - \mathcal{J}^{'}(0) + \mathcal{J}^{'}(\mathbf{Q}) + h_{A})^{2} - |\mathcal{J}(\mathbf{Q})|^{2}},
\label{eq:omega}
\end{equation}
where S is spin and one defines a reduced anisotropy field with units of energy:
\begin{equation}
h_{A} = \dfrac{g\mu_{B}H_{A}}{2S}.
\end{equation}

Spin-wave theory further predicts that the one-magnon cross-section contains the terms of the form:\\
\begin{equation}
\dfrac{d^{2}\sigma}{d\Omega dE^{'}}(\bm{\kappa},\omega) \propto \\
 \sum_{\mathbf{q},\bm{\tau}} n_{\mathbf{q}} \delta(\hbar \omega_{\mathbf{q}} - \hbar \omega) \delta(\bm{\kappa}-\mathbf{q}-\bm{\tau}) \{ u_{\mathbf{q}}^{2} + v_{\mathbf{q}}^{2} +
2u_{\mathbf{q}}v_{\mathbf{q}}\cos(\bm{\rho}\cdot\bm{\tau})\},
\end{equation}
where $\bm{\tau}$ are reciprocal lattice vectors for a single sublattice and $\bm{\rho}$ is the vector connecting opposite sublattices. For the diamond lattice, $\bm{\tau}$ are the reciprocal lattice vectors for the FCC lattice ($h,k,l$ all even or all odd) and $\bm{\rho} = \frac{a}{4}(1~ 1~ 1)$.

The functions $u_{\mathbf{q}}$ and $v_{\mathbf{q}}$ are determined from the equations:\\
\begin{eqnarray}
u_{\mathbf{q}}^{2} &=& \dfrac{S(\hbar\omega_{\mathbf{q}} + 2S \mathcal{J}(0) - 2S\mathcal{J}^{'}(0) + 2S\mathcal{J}^{'}(\mathbf{Q}) + g\mu_{B}H_{A})}{\hbar\omega_{\mathbf{q}}}\\
u_{\mathbf{q}}v_{\mathbf{q}} &=& \dfrac{-2S^{2}\mathcal{J}(\mathbf{q})}{\hbar\omega_{\mathbf{q}}}
\end{eqnarray}

Equations (2)-(8) predict that spin-wave excitations out of a N$\mathrm{\acute{e}}$el ordered state on a diamond lattice will have band minima at every reciprocal space position where the selection rules for the FCC lattice are satisfied, with a gap at the magnetic zone center determined by the size of $h_{A}$. Intensity is seen to be enhanced at wavevectors where only diamond-lattice magnetic peaks are predicted, suppressed at those wavevectors where only nuclear peaks are predicted and remain at intermediate values where both are observed. The above formula were used to fit the observed dispersion of the magnon mode in $\mathrm{CoAl_{2}O_{4}}$, and the results are depicted by the red curves in Fig. (4) of the main article.

\section{Fit Function for Time-of-Flight Spectra}

To quantify the temperature dependence of the scattering at the magnetic zone center, cuts through the cold-neutron time-of-flight were fit to a sum of elastic and inelastic terms. The elastic scattering from the presence of the magnetic Bragg peak was approximated by a resolution limited Bragg peak:

\begin{equation}
G(\omega) = \frac{A_{1}}{\sqrt{2\pi}\sigma}e^{-\frac{\omega^{2}}{2\sigma^{2}}},
\end{equation}

where $\sigma$ is the resolution width in energy.

The inelastic scattering terms were dictated by the fluctuation-dissipation theorem as it applies to neutron scattering\cite{shirane_book}:

\begin{equation}
S(\bm{Q},\omega) = \dfrac{\chi^{\prime\prime}(\bm{Q},\omega)}{1-e^{-\hbar \omega /k_{B}T}},
\end{equation}

where $S(\bm{Q},\omega)$ is the scattering function and $\chi^{\prime\prime}(\bm{Q},\omega)$ is the imaginary part of the generalized magnetic susceptibility.

In particular, inelastic scattering was modeled as a sum of a damped-harmonic-oscillator term

\begin{equation}
DHO(\omega) = \frac{A_{3}\Gamma_{3}}{\pi \omega^{'}} \left(\dfrac{1}{\Gamma_{3}^{2}+(\omega-\omega^{'})^{2}}-\dfrac{1}{\Gamma_{3}^{2}+(\omega+\omega^{'})^{2}}\right) \dfrac{1}{1-e^{-\hbar \omega /k_{B}T}},
\end{equation}

\begin{equation}
\omega^{'2}= \omega^{2} - \Gamma_{3}^{2}
\end{equation}

to account for spin-wave excitations out of the N$\mathrm{\acute{e}}$el ordered state, and a Lorentzian term

\begin{equation}
L(\omega) = \dfrac{A_{2}\Gamma_{2}}{\pi}\frac{\omega}{(\Gamma_{2}^{2}+\omega^{2})}\dfrac{1}{1-e^{-\hbar \omega /k_{B}T}}
\end{equation}

to account for the observed quasi-elastic scattering, presumably arising from a disorder fraction spins. The results of fits of this kind are described in the main article.

\clearpage
\section{Figure Legends}
\vspace{10mm}
\noindent
\textbf{Figure S1}

Measurements of (\textbf{a}) magnetization and (\textbf{b}) heat capacity as a function of temperature on samples from this study.

\vspace{10mm}
\noindent
\textbf{Figure S2}

A representative rocking curve for a single-crystal used in this study.

\vspace{10mm}
\noindent
\textbf{Figure S3}

A comparison of radial scans across the (0 0 2) reciprocal lattice position at T = 25 K with the crystal aligned in two different scattering planes.

\vspace{10mm}
\noindent
\textbf{Figure S4}

A comparison of fitted $\kappa$ parameters from elastic neutron scattering data on two A-site spinel materials: $\mathrm{CoAl_{2}O_{4}}$ and $\mathrm{MnAl_{2}O_{4}}$. Temperature is given in units of magnetic ordering temperature, $T_{c}$, and an overall scaling factor is applied to data for $\mathrm{MnAl_{2}O_{4}}$.

\clearpage
\section{Figures}

\begin{figure}[h]
\centering
\includegraphics[width=\columnwidth]{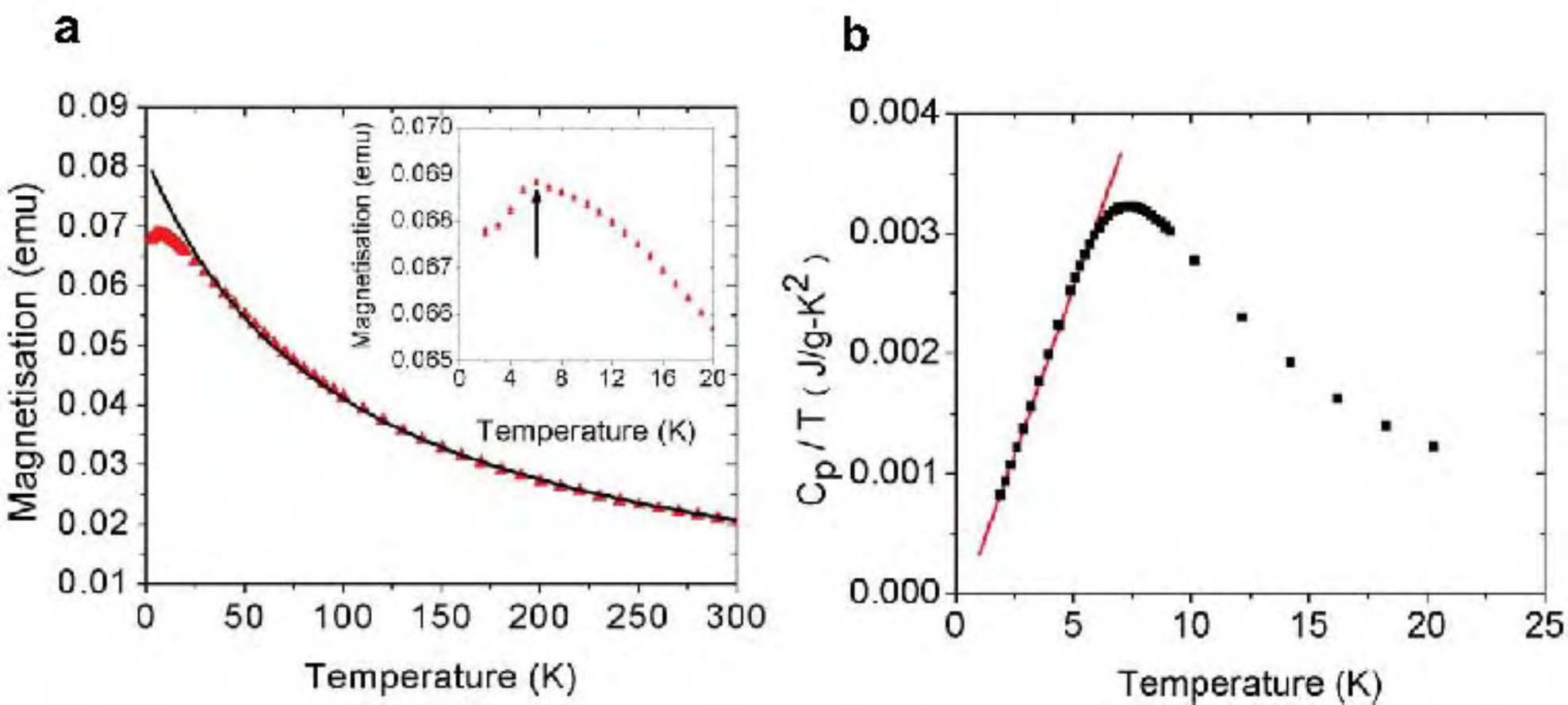}
\caption{\label{fig:structure} }
\end{figure}

\begin{figure}[tpb]
\centering
\includegraphics[width=0.5\columnwidth]{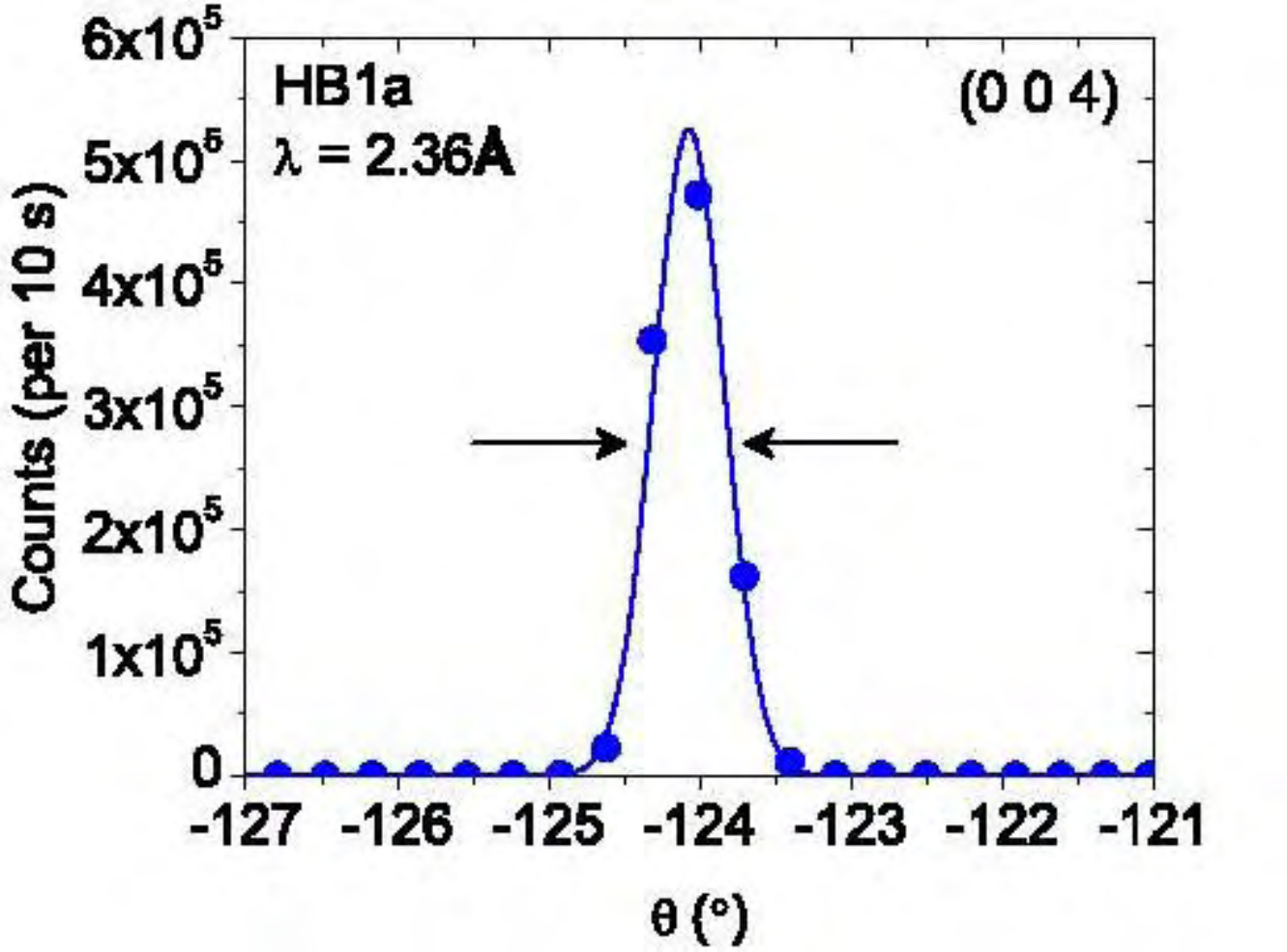}
\caption{\label{fig:TA_diffuse} }
\end{figure}

\begin{figure}[tpb]
\centering
\includegraphics[width=0.5\columnwidth]{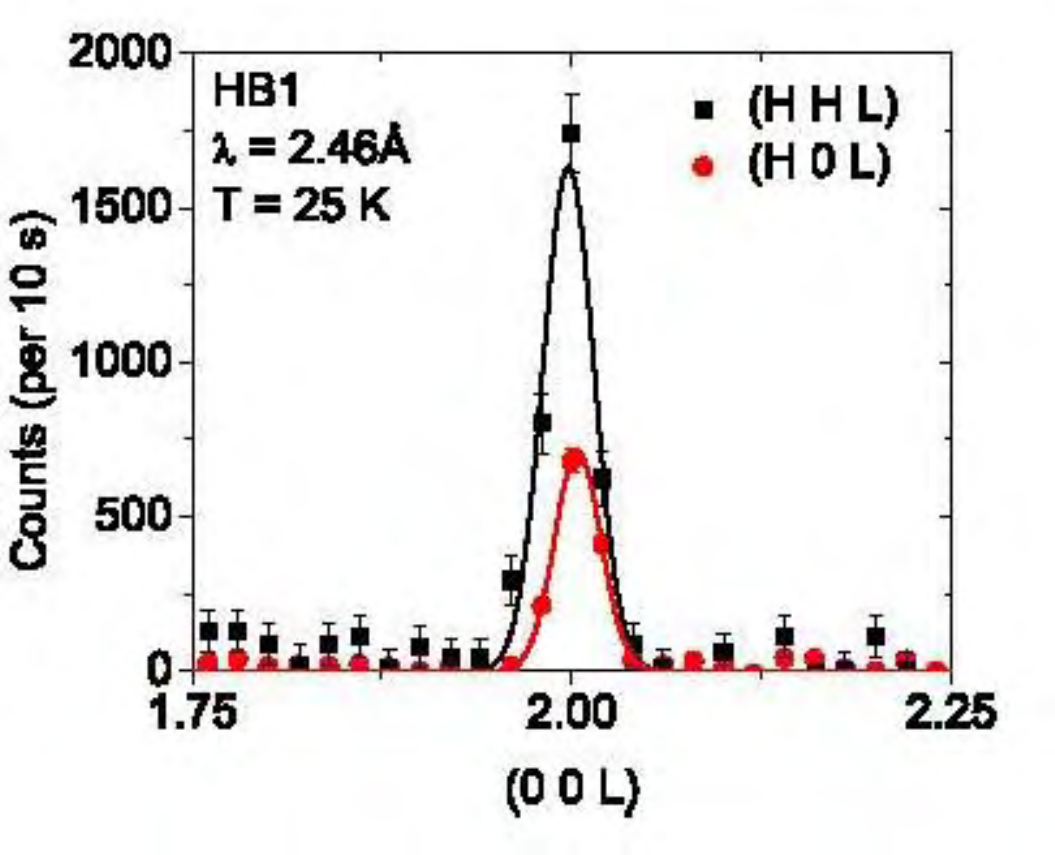}
\caption{\label{fig:TA_fits} }
\end{figure}

\begin{figure}[tpb]
\centering
\includegraphics[width=\columnwidth]{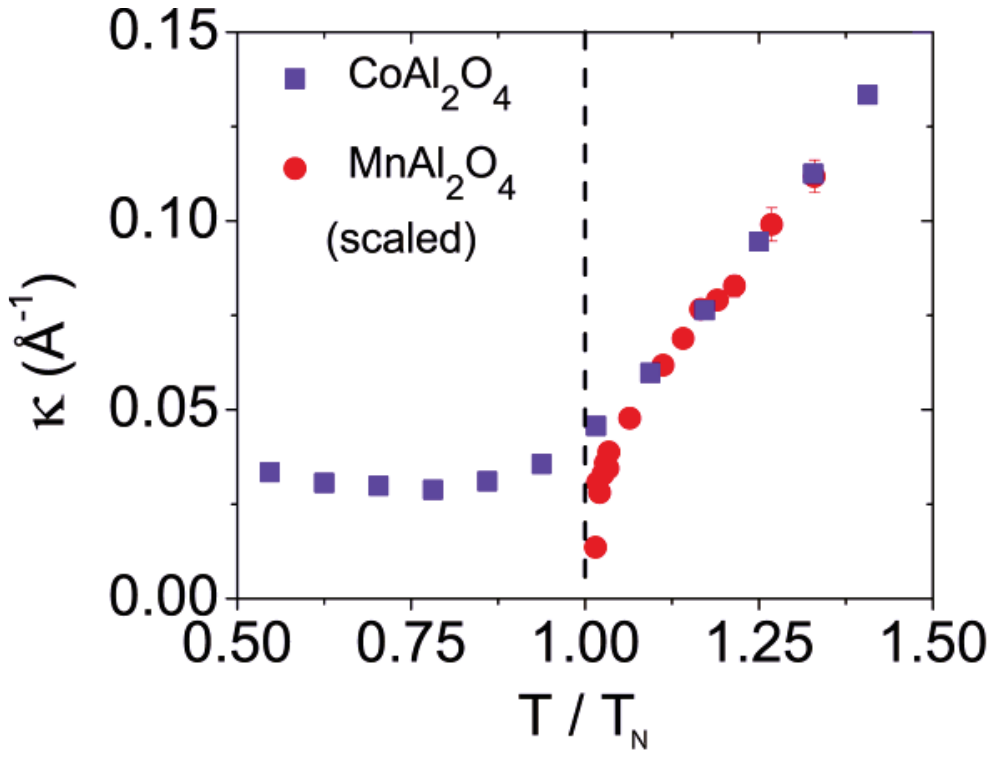}
\caption{\label{fig:TA_inelastic} }
\end{figure}

\clearpage
\section{Tables}
\vspace{10mm}

\textbf{Table S1.} ~Values of reduced $\chi^{2}$ from fits of elastic scattering data to isotropic (I) and anisotropic (A) combinations of Lorentzians (L), Lorentzians squared ($L^{2}$), Lorentzians to an arbitrary power ($L^{p}$) and Gaussians (G).

\begin{table}[h]
\begin{tabular}{|c|c|c|}
	\hline
Fit function & \multicolumn{2}{|c|}{Reduced $\chi^{2}$} \\ \cline{2-3}
 & 2 K & 3 K \\
    \hline \hline
$IG$   & 236 & 208\\ \hline
$AG$   & 224 & 201\\ \hline
$IL^{2}$ & 42.2 & 43.3\\ \hline
$AL^{2}$   & 29.1 & 33.6\\ \hline
$IL$   & 27.4 & 19.3\\ \hline
$AL$   & 20.1 & 13.7\\ \hline
$IL^p$ & 13.5 & 11.3\\ \hline
$IL$ + $IG$  & 16.1  & 12.8\\\hline
$IL$ + $IL^{2}$  & 13.1 & 10.6\\ \hline
$AL$ + $IG$  & 7.56  & 6.63\\ \hline
$IL$ + $AG$  & 6.10  & 4.90\\ \hline
$AL$ + $AG$  & 5.46  & 4.45\\ \hline
$AL^p$ & 4.71 & 3.74\\ \hline
$AL$ + $IL^{2}$  & 4.11  & 4.11\\ \hline
$IL$ + $AL^{2}$  & 1.84 & 2.02\\ \hline
$AL$ + $AL^{2}$  & 1.66  & 2.00\\ \hline
\end{tabular}
\end{table}

\textbf{Table S2.}
Estimations of the number of broken bonds and resultant energy per unit area for domain walls oriented perpendicular to the (0 0 1) and (1 1 0) directions.

\begin{table}[h]
\begin{tabular}{|c|c|c|c|}
	\hline
Wall normal & $N_{broken,1}$ per unit area & $N_{broken,2}$ per unit area &  $E_{wall}$ per unit area  \\
	\hline
(0~ 0~ 1) & 4.0 & 16.0 & 2.03\\
(1~ 1~ 0) & 2.83 & 16.9 & 0.94\\
     \hline
\end{tabular}
\end{table}

\end{document}